\documentclass[10pt]{iopart}
\usepackage{epsfig}

    \newcommand{\pathnow}{}
\begin{document}

\title[QGP fireball explosion]{QGP fireball explosion}

\author{J. Letessier$^\dag$,
G. Torrieri$^\ddag$, S. Hamieh$^\dag$ and  J. Rafelski$^\ddag$}

\address{
\dag\  LPTHE,
Universit\'e Paris 7, 2 place Jussieu, F--75251 Cedex 05\\
\ddag\ Department of Physics, 
University of Arizona, Tucson, AZ 85721, USA}

\begin{abstract}
We identify the major physics milestones in the development of
strange hadrons as an observable for both the formation of quark-gluon 
plasma, and of the
ensuing explosive disintegration of deconfined matter fireball formed in 
relativistic heavy ion collisions at 160--20$A$ GeV. We describe the 
physical properties of QGP phase and show  agreement with  the expectations
based on an analysis of  hadron abundances. We than also demonstrate that 
the $m_\bot$ shape of hadron spectra is in qualitative agreement with 
the sudden breakup of a supercooled QGP fireball.
\end{abstract}

\pacs{12.38.Mh}

\submitto{\JPG}

\section{Why Quark-Gluon Plasma?}
\subsection{Strangeness Enhancement}
Considering that 
strangeness and entropy (hadron multiplicity) excess due to QGP 
formation accompany  each other, the appropriate way to measure the global
strangeness enhancement is to study the yield of strangeness produced 
per participant baryon. This can be done in many different ways, one 
often invoked is  to compare proton induced reactions with nucleus-nucleus 
reactions \cite{WA97Fini}. 

The enhanced strangeness yield observed at 158--200$A$ GeV reactions
corresponds according to our study  to  $\cal O$(1) $s\bar s$-pairs
of quarks per participant baryon \cite{Let00}. This exceptionally high yield 
is achievable in a short time that the collision is known to last 
by in-plasma gluon-fusion reactions, $G+G\to s+\bar s$ \cite{Raf82}. 
Strangeness enhancement alone is a possible QGP signature. However, it is hard
to argue that strangeness alone suffices to show formation of QGP phase, as 
other mechanisms of flavor formation within some other new physics scenarios
can be proposed. Quite early on strange antibaryons have been recognized as being more 
specific of the deconfined phase (see next section) \cite{firstS}. 

However, systematic
study of strangeness enhancement offers a relatively simple and 
powerful tool to find the critical behavior of elementary matter. 
At low energies the energy threshold behavior of strangeness formation 
leads to a rapid rise in  strangeness yield. As long as onset of
deconfinement does not produce excess pions, strangeness per pion yield can
appear  at lower collision energies to be greater than it is in QGP phase 
reached at high energies \cite{Gaz00}. 

A comparison of AGS  results with CERN  higher energy  strangeness per
pion yield suggests that the  lower collision energy yield is higher \cite{DG00},
as is in fact consistent with the formation of the QGP phase at higher energies.  
This production pattern of strangeness 
allows a simple determination of the transition energy to 
the high entropy phase, presumably QGP, which is the energy at 
which  the strangeness to pion ratio is  diluted by melting of color
degrees of freedom -- this at first enhances the hadron yield, and 
only when in-plasma temperature and plasma lifespan at higher collision
energy is sufficiently high, the strangeness yield picks up, regaining 
some lost ground in comparison to the pion yield (see Fig. 1 in \cite{Gaz00}).

\subsection{Multistrange Antibaryon Enhancement}
Even though the strange pattern of strangeness excess as function of energy 
would clearly indicate formation of a new state of elementary matter, more
specific signatures are needed to establish the deconfined nature  of the 
new phase. 

Strange antibaryons allow
to probe  mobility and density  of individual strange quarks \cite{firstS}
at the time these otherwise rarely produced hadrons (low backgrounds) are
formed. In QGP plasma  the predicted  \cite{firstS}, and observed  \cite{WA97Fini}, 
enhancement of strange antibaryons occurs with an unusual pattern: enhancement is  
increasing with strangeness content, consistent with these particles being 
emitted  from a very dense source with (nearly) completely saturated 
strangeness phase space \cite{KMR86}. Here, enhancement is defined 
with a base being the expectation derived from scaled p-N reactions. 
In models employing cascading interactions
to create multistrange particles such a pattern is highly unnatural.

A special feature emerges from the study of strange antibaryons at CERN-SPS: they
seem to emerge directly, and without reinteraction in any baryon-rich medium, 
from the strangeness symmetric deconfined phase.
This has led to the suggestion of explosive disintegration of the QGP fireball
which we now briefly discuss.

\section{Why Explosion?}
\subsection{Relative Chemical Equilibrium and Absolute Non-Equilibrium}
The abundance analysis  of relative strange particle abundances 
shows that they follow  a pattern  arising when  the phase space of 
strangeness is symmetric between strange and antistrange quarks. 
This is surprising  in a baryon-rich environment.
A solution of this mystery calls for both strange and
antistrange quarks to  be mobile and not bound in hadronic
 states \cite{Raf91}. Analysis of lower energy 10--14$A$ GeV
AGS results clearly shows the expected matter-antimatter 
asymmetry \cite{RD94}. This `relative chemical equilibrium' study 
suggests that there is  an important difference in the reaction mechanism
arising comparing reactions occurring at CERN and AGS energies. 
The nature of relative chemical equilibrium at CERN energies allows thus to 
suppose that the source has deconfined strange quarks, and that 
the observed strange antibaryons did not rescatter after formation, which is
confirmed by the properties of their $m_\bot$-spectra (see below). 

Analysis of hadron abundances (absolute chemical equilibrium) 
shows that hadrons (in particular strange) are described much more 
precisely when the hadron phase space is evaluated  allowing for 
nonequilibrium quark pair yield \cite{LRPad99}. This means that a 
reaction picture should invoke formation of hadrons  on a global
time scale which is shorter than chemical equilibration for any type of matter
considered.

Further and direct evidence for  
rapid hadron formation comes from HBT correlations analysis, which
 shows that  pions arise within a period of time
compatible with zero, and at most 2fm/c long. Especially for pions one 
would expect that continuous emission from an evolving source, thus 
this result is somewhat unexpected and indicative of a violent end
of the fireball of dense matter \cite{Ste99}.

It can be imagined that the dense mater fireball formed at CERN 
breaks up as fast as causality allows: when the sudden hadronization begins, 
the freeze-out surface propagates at light velocity $c$. However, the flow
velocity we find is  smaller, consistent with velocity of 
sound of relativistic matter $(c/\sqrt{3})$.

\subsection{$m_\bot$-spectra}
Strange antibaryons, specifically 
$\overline\Lambda$ and $\overline\Xi$, $m_\bot$-spectra measured by the experiment 
WA97  at central rapidity averaged over 30\% centrality cannot be distinguished 
from the corresponding shape of $\Lambda$ and $\Xi$ spectra. This contradicts
the behavior expected should strange  antibaryons be evolving when 
embedded in a baryon  rich hadron medium. Furthermore, should the 
production of strange antibaryons occur in hadron gas medium, there 
is no reason to expect that these spectra could be equal, as 
in this scenario fundamentally different processes contribute to the formation of 
particles and antiparticles \cite{KMR86}.

The $m_\bot$-spectra of single and double 
strange baryons and antibaryons are also found to be 
nearly identical.
This effect sets low limits on  formation mechanism involving a gradual build-up 
due to multi-meson reactions, or continuous emission from a 
source that is evolving, and not decoupled from these particles. 

On the other hand, explosive hadronization picture requires 
that the thermal and chemical freeze-out conditions of all hadrons 
(nearly) agree, and that the thermal
freeze-out conditions  of very different particles are also nearly the same.
This is an extraordinarily strong requirement. Thus in order 
to falsify explosive hadronization picture we either:\\
1) show that, {\it e.g.}, pion spectra are inconsistent with
hyperon spectra in an analysis that applies same methodology 
including in the thermal freeze-out analysis 
both temperature and transverse explosion
velocity or/and\\
2) show that $m_\bot$-spectra cannot be described by statistical 
parameters obtained from the chemical freeze-out analysis. 

Such a evaluation  and comparison of freeze-out conditions 
involving both chemical and thermal properties of hadronic 
particles  has not  been previously 
performed in a fully consistent 
fashion, and we will present here the very first results. We find
a good agreement with sudden hadronization scenario.

\section{Mechanism of Sudden Hadronization} 
\subsection{Chemical Freeze-out}
We reported at the last strangeness meeting in Padova
that  in order to arrive at a statistically significant
description of experimental strange hadron abundance results 
we need to introduce  valance quark pair abundance \cite{LRPad99},
as expressed by the parameters $\gamma_q$ and $\gamma_q/\gamma_s\simeq 1$. 
Our ongoing study of the Pb--Pb system final state hadron 
abundances strongly favors  $\gamma_q\simeq \exp (m_\pi/2T)>1, 
\gamma_s\simeq \gamma_i$  \cite{Let00}, confirming and extending 
the ideas expressed in Padova.

We have now also understood that chemical non-equilibrium arises 
naturally in explosive fireball breakup.
The microscopic processes governing the fireball
breakup determine how the physical and statistical properties
of the fireball change at the breakup point. 
The energy $E$  and baryon content  $b$ of the fireball are  conserved. 
Entropy $S$ is conserved when the gluon content of a QGP fireball is transformed into 
quark pairs in the entropy conserving  process $G+G\to q+\bar q$. Similarly, when
quarks and antiquarks recombine into hadrons,  entropy is conserved in the 
range of parameters of interest here.  Thus  also $E/b$ and $S/b$ is conserved 
across hadronization condition. The sudden hadronization process 
also maintains the temperature $T$ and baryo-chemical potential $\mu_b$
across the phase boundary. What changes are the chemical occupancy parameters. 
As gluons convert into quark pairs and hadrons $\gamma_g\to 0$ but
the number occupancy of light valance quark  pairs 
increases $\gamma_q>\gamma_{q_0}\simeq 1$
 increases significantly, along with  the number occupancy 
of strange quark  pairs $\gamma_s>\gamma_{s_0}\simeq 1$.

\subsection{Mechanical Instability and Deep Supercooling}
We found that sudden breakup (hadronization) 
into final state  particles occurs as the fireball super-cools, and in this 
state encounters a strong mechanical instability \cite{Raf00}.
Deep supercooling requires a  first order phase transition, and
this in turn implies presence of a latent heat.  The total pressure 
and energy  comprise particle (subscript $\mbox{\scriptsize p}$)
and the  vacuum properties:
\begin{equation}\label{EPB1}
P^{(i)}=P_{\mbox{\scriptsize p}}-{\cal B}\,,\quad \varepsilon
 =\varepsilon^{(i)}_{\mbox{\scriptsize p}}+{\cal B}\,.
\end{equation} 
The upper index  $(i)$ refers for the intrinsic energy density $\varepsilon$ and
pressure $P$  of matter in the  frame of reference,  
locally at rest, {\it i.e.}, observed by a co-moving observer.  
We  omit the superscript  $(i)$ in the following.

The surface normal vector of exploding fireball is $\vec n$, and the local velocity 
of matter flow $\vec v_{\mbox{\scriptsize c}}$. 
The rate of  momentum flow vector $\vec {\cal P}$ 
at the surface  is obtained from the energy-stress tensor $T_{kl}$ \cite{LanHyd}: 
\begin{equation}\label{Peqv}
\vec {\cal P}=P\vec n+(P+\varepsilon)
  \frac{\vec v_{\mbox{\scriptsize c}}\, \vec v_{\mbox{\scriptsize c}}\!\cdot\! \vec n}
          {1-\vec v_{\mbox{\scriptsize c}}^{\,2}}\,.
\end{equation}

 For the fireball expansion to continue, ${\cal P}\equiv |\vec {\cal P} |> 0$
is required.  For ${\cal P}\to 0$ at $v_{\mbox{\scriptsize c}}\ne 0$, we have
a conflict between the desire of the motion to stop or even reverse, and 
the continued inertial expansion. Expansion beyond  ${\cal P}\to 0$ is in general not possible. 
A surface region of the fireball that  reached it but continues to flow outwards 
must   be torn apart. This is a collective instability and thus the 
ensuing disintegration of the fireball matter will be 
very rapid, provided that much of the surface reaches this condition.
We adopt the condition $\vec {\cal P}=0$ at any surface region to be
the  instability condition of an expanding hadron 
matter fireball.

\section{QGP fireball}
\subsection{Equations of State}
In order to quantitatively evaluate  supercooling of a QGP fireball,
we need a model of equations of state of the deconfined phase that 
comprises interactions, and specifically, the freezing of the color
degrees of freedom, as temperature decreases. We have developed a model which
agrees well with the lattice results and have presented it elsewhere \cite{Ham00}.
The key ingredients of our model are:
\begin{enumerate}
\item
We relate the QCD scale to the temperature $T=1/\beta$, 
 we use for the scale the Matsubara frequency \cite{Pes00}
(we are following the notation convention and thus $\mu$ 
without a subscript is {\it not} a chemical potential, it is
the QCD scale):
\begin{equation}
\label{runalTmu}
\mu=2\pi \beta^{-1}\sqrt{1+\frac{1}{\pi^2}\ln^2\lambda_{\mathrm q}}
=2\sqrt{(\pi T)^2+\mu_{\mathrm q}^2}\,.
\end{equation}
This extension to finite chemical  potential $\mu_{\mathrm q}$, or 
equivalently quark fugacity  $\lambda_{\mathrm q}=\exp{\mu_{\mathrm q}/T}$, 
is motivated by the form of plasma frequency  entering
the computation of the vacuum polarization function \cite{Vij95}.
\item
We obtain the interacting strength $\alpha_s(\mu)$
integrating numerically the renormalization group equations 
\begin{equation}
\label{dalfa2loop}
\mu \frac{\partial \alpha_s}{\partial \mu}=
-b_0\alpha_s^2-b_1\alpha_s^3+\ldots \equiv \beta^{\mbox{\scriptsize pert}}_2\,.
\end{equation}
$\beta^{\mbox{\scriptsize pert}}_2$ is
the beta-function of the renormalization group 
in two loop approximation, and 
$$b_0=\frac{11-2n_{\mathrm f}/3}{2\pi}\,,\quad 
   b_1=\frac{51-19n_{\mathrm f}/3}{4\pi^2}\,.$$ 
$\beta^{\mbox{\scriptsize pert}}_2$
does not depend on the renormalization scheme,
and solutions of Eq.\,(\ref{dalfa2loop}) differ from higher 
order renormalization scheme
dependent results  by less than the error introduced by the experimental 
uncertainty in the measured value of $\alpha_s(\mu=M_Z)=0.118+0.001-0.0016$. 
\item
We introduce, in the domain of freely mobile quarks 
and gluons, a finite vacuum energy  density:
$${\cal B}=0.19\,\frac{\mbox{GeV}}{\mbox{fm}^3}\,.$$
This also implies, by virtue of relativistic invariance,
that there must be a (negative) 
associated pressure acting on the surface of this volume, 
aiming to reduce the size of the deconfined region.  
These two properties of the vacuum follow
consistently from the vacuum partition function:
\begin{equation}
\label{Zbag}
\ln{\cal Z}_{\mbox{\scriptsize vac}}\equiv -{\cal B}V\beta\,.
\end{equation}
\item 
The partition function of the quark-gluon liquid comprises interacting 
gluons, $n_{\mathrm q}$ flavors of light quarks  \cite{Chi78}, 
and the vacuum ${\cal B}$-term. We
incorporate further the strange quarks by assuming that their mass 
in effect reduces their effective number  $n_{\mathrm s}<1$:
\begin{eqnarray}
\label{ZQGPL}
&&\frac{T}{V}\ln{\cal Z}_{\mathrm QGP}
\equiv P_{\mathrm QGP}=
-{\cal B}+\frac{8}{45\pi^2}c_1(\pi T)^4  \nonumber\\ 
&&+
\frac{n_{\mathrm q}}{15\pi^2}
\left[\frac{7}{4}c_2(\pi T)^4+\frac{15}{2}c_3\left(
\mu_{\mathrm q}^2(\pi T)^2 + \frac{1}{2}\mu_{\mathrm q}^4
\right)\right]
\nonumber\\ 
&&+
\frac{n_{\mathrm s}}{15\pi^2}
\left[\frac{7}{4}c_2(\pi T)^4+\frac{15}{2}c_3\left(
\mu_{\mathrm s}^2(\pi T)^2 + \frac{1}{2}\mu_{\mathrm s}^4
\right)\right]\,,
\end{eqnarray}
 where:
\begin{eqnarray}
\label{ICZQGP}
c_1&=&1-\frac{15\alpha_s}{4\pi}+ \cdots\,,\\ \nonumber
c_2&=&1-\frac{50\alpha_s}{21\pi}+ \cdots\,,\qquad
c_3=1-\frac{2\alpha_s}{\pi}+ \cdots\,.
\end{eqnarray}
\end{enumerate}

\subsection{Properties of QGP}
\begin{figure}[bt]
\vspace*{0.cm}
\centerline{\hskip 0.5cm
\hspace*{5.0cm}\epsfig{width=12.2cm,clip=,figure=\pathnow 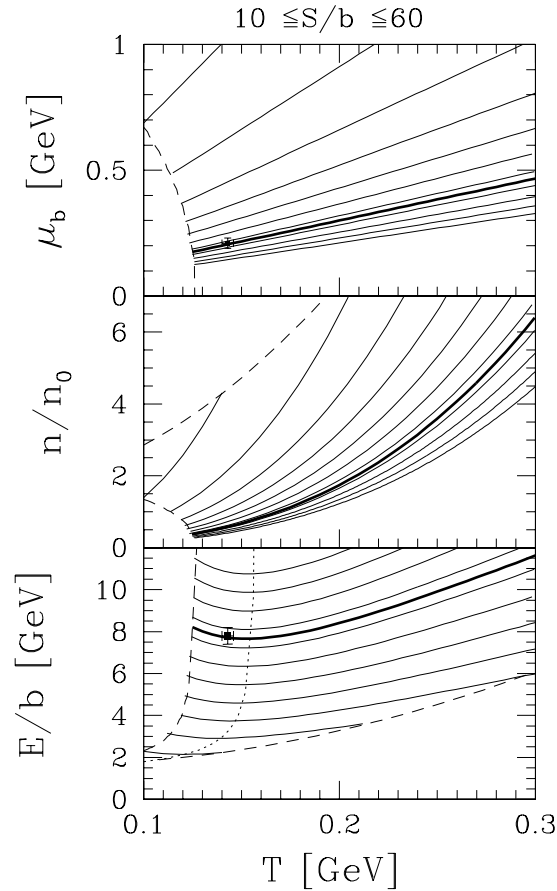}
}
\vspace*{-0.6cm}
\caption{ 
 From top to bottom: $\mu_b,\ n/n_0$ and $E/b$; 
lines shown correspond to fixed entropy per baryon 
$S/b= 10$ to 60 by step of 5 (left to right). 
Thick solid lines: result for $S/b=42.5$.
Limits:  energy density $\varepsilon_{\mathrm q,g}=0.5$\,GeV/fm$^3$ and
baryo-chemical potential $\mu_b=1$\,GeV. The experimental points denote
chemical freeze-out analysis result \protect\cite{Let00}.
\label{EOSSfix}
}
\end{figure}

We show properties of the quark-gluon liquid in a wider range of parameters
at fixed entropy per baryon $S/b$ in the range $S/b=10$--60 
 in step of 5 units. In the  top panel  in 
figure \ref{EOSSfix}, we show  
baryo-chemical potential $\mu_b$, in middle panel
baryon density $n/n_0$, here  $n_0=0.16/\mbox{fm}^3$, and bottom left
the energy per baryon $E/b$. In top and middle 
 panel the low entropy results are top-left in figure, in bottom panel bottom left. 
The  highlighted curve, in figure \ref{EOSSfix},
is for the  value $S/b=42.5$\,. The dotted line, at the 
minimum of $E/b\vert_{S/b}$, is where the vacuum and quark-gluon gas pressure balance. 
This is the equilibrium point and indeed  the energy per 
baryon does have a relative minimum there. 

Since little entropy is produced during the evolution of 
the QGP fireball, lines in the lower panel of figure \ref{EOSSfix}
characterize the approximate trajectory in time of the fireball. After
initial drop in energy per baryon due to transfer of energy to accelerating 
expansion of the fireball, during the deep supercooling process the motion 
is slowed and thus energy per baryon increases. However only at the entropy 
10 the two processes balance and the collective motion slows. At higher entropy
content the vacuum pressure is not sufficient to stop the explosion of a 
fireball. The thick line is our expectation for the fireball 
made in Pb--Pb interactions at the projectile energy 158$A$ GeV. 
The cross shows the result of  chemical freeze-out 
analysis. 

\begin{figure}[tb]
\vspace*{-2cm}
\centerline{\hskip 0.5cm
\hspace*{1.0cm}\epsfig{width=10.5cm,clip=,figure=\pathnow 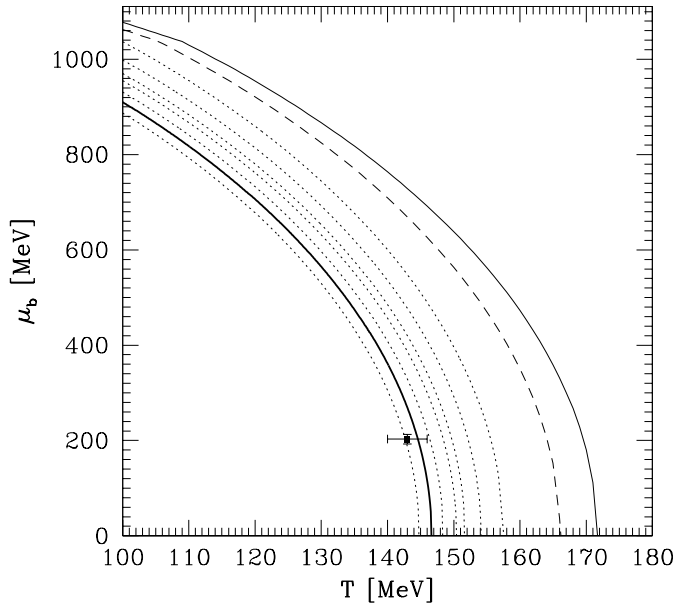}
}
\vspace*{-0.8cm}
\caption{ 
Thin solid and dashed lines: equilibrium phase transition from
hadron gas to QGP liquid without and with excluded volume correction,
respectively.  Dotted:  
breakup condition at shape parameter $\kappa=0.6$, for expansion velocity 
$v_{\mbox{\scriptsize c}}^2=0, 1/10, 1/6, 1/5, 1/4$ and 1/3, and thick line for 
$v_{\mbox{\scriptsize c}}=0.54$. The experimental point denotes 
chemical non-equilibrium freeze-out analysis result~\protect\cite{Let00}.
\label{PLMUPLIQ}
}
\end{figure}

We also have compared the chemical freeze-out conditions with the
phase  transition properties. 
The thin solid line in the $T,\mu_{\mbox{\scriptsize b}}$ plane  
in figure \ref{PLMUPLIQ} shows  where the pressure of the quark-gluon
liquid  equals the equilibrated hadron gas pressure. 
The hadron gas behavior is obtained evaluating and summing 
the contributions  of  all known hadronic resonances considered to be
point particles. When we allow for 
finite volume of hadrons \cite{HR80}, we find that the hadron  pressure is
slightly reduced, leading to some (5\,MeV) reduction in the  equilibrium transition 
temperature, as is shown by  the dashed line in figure \ref{PLMUPLIQ}\,.
For vanishing baryo-chemical potential, 
we note in figure \ref{PLMUPLIQ} that the equilibrium
phase transition temperature is  $T_{\mbox{\scriptsize pt}}\simeq 172$\,MeV,
and when finite hadron size is allowed, $T_{\mbox{\scriptsize fp}}\simeq 166$\,MeV,
The scale in temperature we discuss is result of 
comparison with lattice gauge results. 
Within the lattice calculations \cite{Kar00}, 
it arises from the comparison with  the string tension.

The dotted lines, in figure \ref{PLMUPLIQ}, correspond to 
break up velocity  for (from right to left) 
$v_{\mbox{\scriptsize c}}^2=0, 1/10, 1/6, 1/5, 1/4$ and $1/3$. The
last dotted line corresponds thus to an expansion flow with 
the velocity of sound of  relativistic noninteracting massless gas. The thick
solid line corresponds to an expansion with $v_{\mbox{\scriptsize c}}=0.54$.  
The hadron analysis result is also shown \cite{Let00}. 
Comparing in figure \ref{PLMUPLIQ} 
thin solid/dashed with the  thick line,
we recognize the deep supercooling as required for the explosive 
fireball disintegration. The super-cooled  zero pressure $ P=0$ QGP  
temperature is at $T_{\mbox{\scriptsize sc}}=157$\,MeV, 
(see the intercept of the first dashed line to the right 
in figure \ref{PLMUPLIQ}) and an expanding fireball can deeply 
super-cool to $T_{\mbox{\scriptsize dsc}}\simeq 147$\,MeV 
(see the intercept of thick solid line)
before the mechanical instability occurs.

\section{Thermal freeze-out}
As noted earlier on, it is essential for the idea of sudden
freeze-out of hadrons that all spectra are well described by the same  
parameters obtained in chemical analysis of (strange) hadrons produced.
Some spectacular results related to this objective 
are shown in the following. These were obtained taking chemical
freeze-out temperature value for the thermal freeze-out 
temperature: $T_f=143$\,MeV, with  a 
freeze-out surface velocity $0.99c$, as expected for sudden break-up of an
unstable fireball. 

\begin{figure}[tb]
\centerline{\hskip 0.5cm
\hspace*{1.0cm}\epsfig{width=8.8cm,clip=,figure=\pathnow 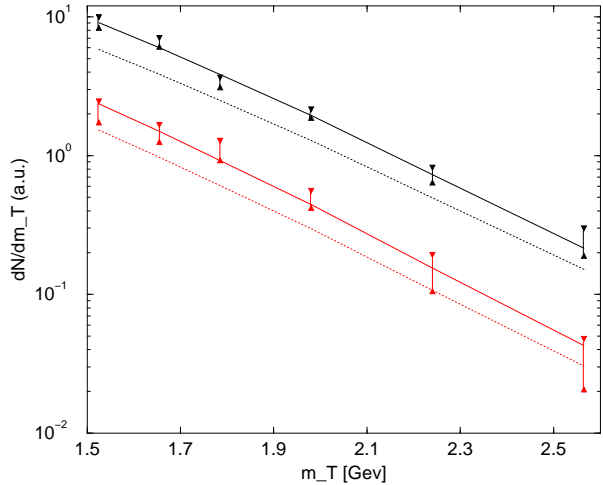}
}
\vspace*{-0.8cm}
\caption{ 
Central rapidity data of experiment WA97 for $\Xi$ (top) and $\overline\Xi$ are 
compared to the spectra expected in a sudden freeze-out reaction picture.
Dotted: directly produced particles. See text for details.
\label{PaLambda1}
}
\end{figure}

In figure \ref{PaLambda1}, we show how the data of WA97 experiment  for 
$\Xi$ (top) and $\overline\Xi$ are understood. We have included as usual 
the contributions from the decay 
$\overline{\Xi}^*(1530)\to \overline\Xi + \pi$
and similarly for $\Xi$. Given the large statistical errors we did 
not (yet)  include in analysis other resonances. Varying the two
parameters normalization and the flow velocity $v_c$, 
we found that $\overline\Xi$ are best described
by  $v_c=0.50\pm0.03$, while for $\Xi$ we found $v_c=0.49\pm0.05$,
both values are consistent with the chemical analysis result $v_c=0.54\pm0.03$.
The $\chi^2/\mbox{dof}$ for $\overline\Xi$ is 0.6 and it is 1.1 for $\Xi$. These
results are remarkably consistent with the results of chemical 
freeze-out analysis \cite{Let00}, and are highly statistically significant. We thus
can draw the conclusion that the double-strange baryons and antibaryons which 
comprise practically all newly made quarks, are produced predominantly 
in a sudden hadronization  process.

\begin{figure}[tb]
\vspace*{-1.3cm}
\centerline{\hskip 0.5cm
\hspace*{1.0cm}\epsfig{width=8.8cm,clip=,angle=-90,figure=\pathnow 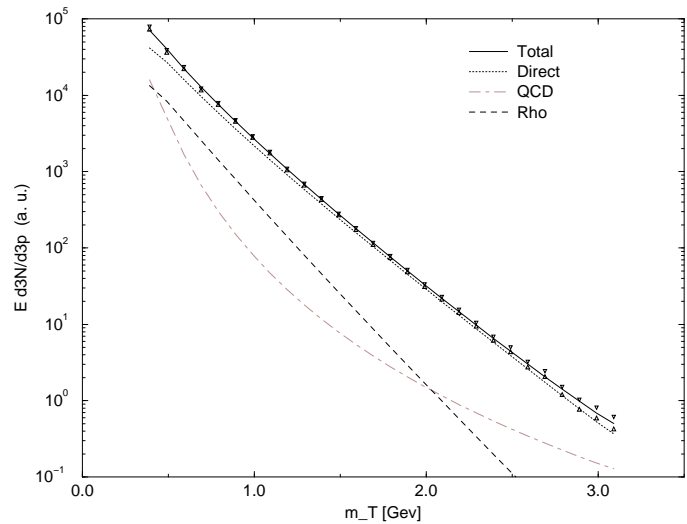}
}
\vspace*{-0.8cm}
\caption{ 
Central rapidity data of experiment WA98 for $\pi^0$ are 
compared to the spectra expected in a sudden freeze-out reaction picture.
See text for details.
\label{Ppi01}
}
\end{figure}

We now turn to the pion spectra. We study the $\pi^0$ data in a very wide
range of $m_\bot$ in which the yield changes
by 6 orders of magnitude.  These results were 
also obtained at central rapidity by the WA98 collaboration \cite{WA98pi0}.
We use the same set of parameters as for cascades, {\it i.e.,} $T_f=143$\,MeV, 
chosen at the chemical freeze-out, and with freeze-out surface velocity $0.99c$. 
We  include in the results shown
in figure \ref{Ppi01} aside of  directly produced pions the two body decay of 
the $\rho$. However, noticing the large range of yield here considered, we  allow 
a direct hard parton QCD component contribution \cite{Fey77},
of the form $Ed^3N/d^3p\propto 1/p_\bot^\kappa$\,. 
Thus, we vary four parameters
two normalizations, and also $\kappa= 5.6\pm1.2$ , $v_c=0.55\pm0.02$. This 
yields again an extremely good description of the data as seen 
in figure \ref{Ppi01}, with $\chi^2/\mbox{dof}<1.4$. Considering that only
statistical error is being  considered, and we were not evaluating 
contributions of 3-body decay resonances, we see this as a highly significant result. 
We note that the hard
scattering contribution is at 1--20\% level, and thus one must make a more 
precise model of this contribution before searching for a precise  description.

Our analysis is not contradicting results shown in \cite{WA98pi0}, for these
authors did not consider that the freeze-out surface velocity is different from 
the flow velocity, and they  did not allow for direct parton-parton scattering
contributions in their analysis of pion spectra.  Moreover, we confirm 
the finding that the apparent temperature hierarchy for different mass
particles  is due to a collective
expansion of the source \cite{NuX98}.

\section{Discussion and Conclusion}
We recall here the conclusion of  Cs\"org\H{o} and Csernai  \cite{Cso94}
who required 
as verification for the presence of a deeply super-cooled state of matter and sudden 
hadronization: 
i) short duration and relatively short mean proper-time of 
particle emission, now seen in particle correlations  \cite{Ste99};
ii) clean strangeness signal of QGP \cite{WA97Fini};
iii) universality of produced particle spectra
which are the remarkable features of strange particle production \cite{WA97Fini};
v) no mass shift of the phi-meson; despite extensive search such a 
 shift has not been found by the NA49 collaboration \cite{NA49phi}.
All this has been found true, and more. 

Is our picture of fireball evolution compelling?
We found that particle production  occurred at condition of 
negative  pressure expected in a deeply 
super-cooled state and have shown 
internal  consistency with (strange) hadron  production
analysis involving chemical non-equilibrium. Moreover, 
these chemical freeze-out conditions agree with
thermal analysis, allowing the conjecture that the 
explosive quark-gluon fireball breakup forms final state hadrons, which do not 
undergo further re-equilibration.

Our first glance at the thermal hadron spectra is clearly
showing  consistency of thermal freeze-out between very different hadrons.
Moreover, we find consistency of thermal and chemical freeze-out conditions.
Both results are required if hadrons are formed in sudden hadronization 
of a supercooled quark-gluon plasma.

\ack
Supported  by a grant from the U.S. Department of
Energy,  DE-FG03-95ER40937\,. Laboratoire de Physique Th\'eorique 
et Hautes Energies, LPTHE, at  University Paris 6 and 7 is supported 
by CNRS as Unit\'e Mixte de Recherche, UMR7589.

\end{document}